\documentclass[letterpaper, 10 pt, conference]{ieeeconf}  

\usepackage{xcolor}
\usepackage{graphicx}
\usepackage{algorithm,algorithmic}
\usepackage{amsmath}
\usepackage{amssymb}
\usepackage{soul}
\IEEEoverridecommandlockouts                              

\begin{document}
\title{\LARGE \bf 
Output-Sampled  
Model Predictive Path Integral Control (o-MPPI)\\ for Increased Efficiency
}

\author{Leon (Liangwu) Yan and Santosh Devasia
\thanks{
Leon (Liangwu) Yan (liangy00@uw.edu) and Santosh Devasia (IEEE Fellow, devasia@uw.edu, see
http://faculty.washington.edu/devasia/) are with the Department of
Mechanical Engineering, University of Washington, Seattle,
}}
\maketitle
\thispagestyle{empty}
\pagestyle{empty}

\begin{abstract}
The success of the model predictive path integral control (MPPI) approach depends on the appropriate selection of the input distribution used for sampling. 
However, it can be challenging to select inputs that satisfy output constraints in dynamic environments. The main contribution of this paper is to propose an output-sampling-based MPPI (o-MPPI), which improves the ability of samples to satisfy output constraints and thereby increases MPPI efficiency. 
Comparative simulations and experiments of dynamic autonomous driving of bots 
around a track are provided to show that the proposed o-MPPI is more efficient and requires substantially (20-times) less number of rollouts and (4-times) smaller prediction horizon when compared with the standard MPPI for similar success rates. The supporting video for the paper can be found at {\sffamily{https://youtu.be/snhlZj3l5CE}}.
\end{abstract}

\section{Introduction}
Sampling-based approaches 
such as model predictive path integral control (MPPI)~\cite{williams2017model} have become popular methods to solve optimization problems due to fast computations possible with graphic processing units and parallelized computing. In such methods, models are used to predict the cost for a series of input rollouts, and the final input selection is a cost-weighted average of the rollouts. An advantage of the sample-based approach is that it is derivative-free, does not require approximation of the system dynamics and cost functions, and allows for non-differentiable cost functions~\cite{mohamed2020model}.

Successful use of the MPPI algorithm requires proper selections of the mean and the covariance of the input samples. If the selected mean leads to a rollout that enters inside an infeasible region (this could be caused by the poor initialization, system disturbance, or sudden environment changes), then most of the sampled rollouts can result in failure~\cite{yin2022trajectory}. Although a larger covariance can be used to alleviate failures by increasing  exploration~\cite{tao2023rrt}, it also often requires a larger number of samples and increased computation load, and can lead to undesirable input chattering~\cite{williams2017model,kim2022smooth}.
Therefore, there is substantial ongoing interest in methods to appropriately select the input samples to improve MPPI performance. 
However, in general, it is challenging to appropriately select a desirable input mean to meet constraints in the output space, especially in dynamic environments. In general, even when the reference input can be selected as desired outputs to a closed-loop system that satisfies constraints, it does not ensure that the output achieves precision tracking of the desired trajectories. 

\begin{figure}[!ht]
\centering
\includegraphics[width=0.8\columnwidth]{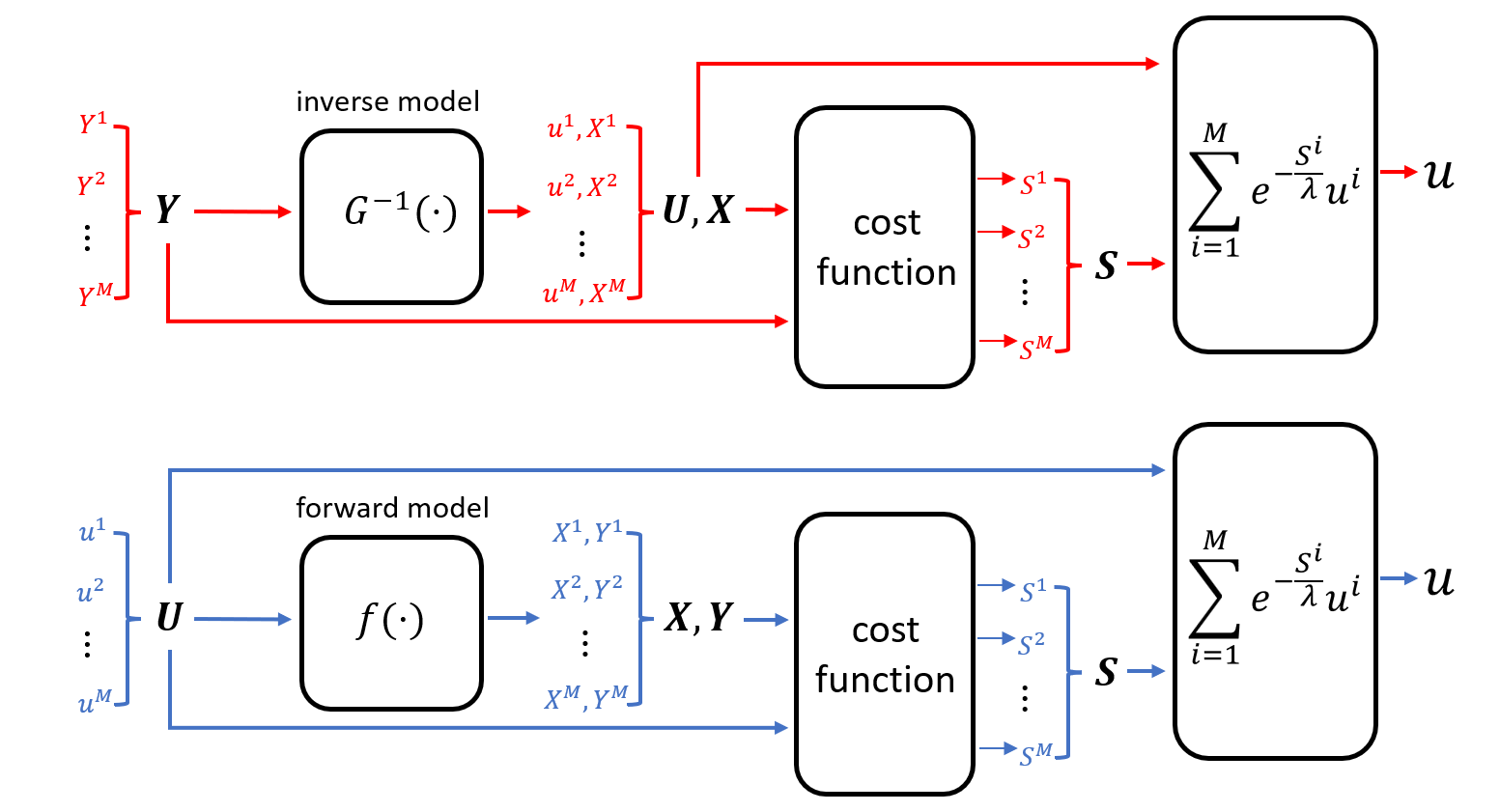}
\vspace{-0.15in}
\caption{Comparison between the proposed output-MPPI (o-MPPI) in red (top) and the standard MPPI in blue (bottom). There are  $M$ rollouts and $f(\cdot)$ maps inputs to states and outputs in the standard MPPI and the inverse  $G^{-1}(\cdot)$ is the inverse dynamics that maps output trajectories to inputs and the states in the proposed o-MPPI.  $S^m$ is the cost for the $m^{th}$  rollout  output-state-input $(Y^m, X^m,u^m)$.
 $\lambda \in R^+$ is the temperature parameter used in the weighting to obtain the optimized input $u$.
}
\vspace{-0.2in}
\label{fig_framework1}
\end{figure}

With the goal of increasing rollouts that satisfy output constraints,  the current paper proposes the sampling of trajectories directly from the output space.  
Then, the proposed output-sampling-based MPPI (o-MPPI) approach uses the inverse model $G^{-1}(\cdot)$ to map outputs to inputs rather than the traditional approach in MPPI of using the input-to-output forward model $f(\cdot)$ as illustrated in  Fig.~\ref{fig_framework1}(top). The inputs obtained using the inverse model are then weighted to obtain the optimized control sequence, as in standard MPPI.  

\vspace{0.001in}
The main contributions of the paper are the following.
\begin{enumerate}
\item 
It proposes an output-sampling-based MPPI (o-MPPI) using inverse models, which improves the ability to select rollouts that satisfy typical output constraints. An advantage of output sampling is that it can leverage well-established trajectory planning algorithms in robotics to select output samples and then use inverse models (physics-based or data-based) to find the associated inputs. This is different from selecting inputs and then finding the associated outputs using forward models.  
\item 
Comparative simulations and experiments of autonomous control of bots (TurtleBot3 Burger)
in a dynamic autonomous driving 
around a track, are used to show that the proposed o-MPPI requires substantially (20-times) less number of rollouts and (4-times) smaller prediction horizon when compared with the standard MPPI to achieve a similar success rate.
\end{enumerate} 

The paper is organized as follows.  Section II discusses the related works. 
Section III describes the proposed o-MPPI followed by its application to an autonomous driving scenario with moving obstacles in Section IV.
Simulation and experimental results are comparatively evaluated and discussed in Section V. Finally, the conclusions and future works are presented in Section VI.

\section{Related work}
\subsection{MPPI with adjusted trajectory distribution}
Several efforts are made to improve the robustness and sample efficiency of standard MPPI by appropriately adjusting the trajectory distribution using different sampling strategies~\cite{yin2022trajectory,bhardwaj2022storm, mohamed2022autonomous,asmar2023model}. 
For example,  warm-starts are used~\cite{williams2017model} to improve  MPPI-variants.
In~\cite{bhardwaj2022storm}, covariances are adapted to accommodate different actions for better exploration. In~\cite{mohamed2022autonomous}, the samples are drawn from a normal and log-normal (NLN) distribution instead of the Gaussian distribution, which can yield improved performance in cluttered environments. 
In~\cite{asmar2023model}, the diagonal covariance matrix $\Sigma_{\epsilon}$ is changed into a nondiagonal covariance matrix and updated through the adaptive importance sampling procedure.
Recently, constraints were added to the terminal state of the prediction horizon in~\cite{yin2022trajectory} to adjust the input distribution. In all these works sampling is in the input space followed by the use of forward models to determine the outputs. In contrast, the current work proposes sampling in the output space (since constraints might be more easily satisfied in the output space) and then uses an inverse model to find the corresponding inputs. Once the input-output pairs are found, the proposed o-MPPI is similar to standard MPPI and can use the same cost functions and weighting strategies. 

\subsection{MPPI with output-space-informed mean}
The output space has been leveraged in the past to improve the input-sampling efficiency of MPPI. 
Since a better mean for the input distribution can also help improve MPPI, an appropriately selected output can be used to inform the mean selection used in standard MPPI. For example, in \cite{tao2023rrt}, the fast rapid-exploring-tree ($RRT^*$) algorithm is used to provide guidance about the mean value of the input for improving sampling efficiency in the input space. In \cite{kusumoto2019informed}, trained Conditional Variational Autoencoders that take into account the contextual information are used to better inform the mean values by taking into the control uncertainties.  Thus, the above methods utilize output-space information to guide the selection of the mean value for the input sampling. The proposed o-MPPI approach extends this idea and fully samples in the output space and then uses inverse models to find the corresponding input?. 
It is noted that sampling of the output space has been used in the path-planning community to smooth and potentially optimize feasible trajectories, e.g.,~\cite{kalakrishnan2011stomp,ma2015efficient}. Here, these planned trajectories are inputs to the closed-loop system, and forward models (of the closed-loop system) are used to predict the system response and for optimizing the selected input (final output trajectory) to the closed-loop system. The proposed o-MPPI can be used with any trajectory planning algorithm such as the fast rapid-exploring-tree ($RRT^*$)~\cite{tao2023rrt} -- the main difference is that inverse models are used in o-MPPI to find inputs that track the selected output trajectories. 

\subsection{Inverse models}
\label{subsec_inv_model_options}
The proposed o-MPPI leverages the strong history in inverse dynamics to find maps from outputs to inputs. 
For example, physics-based models can be used to analytically find the inverse input~\cite{devasia1996nonlinear} in robotics applications. When the physics-based models are not sufficiently precise, the Gaussian process can be used to learn the discrepancy~\cite{meier2016towards}. The entire inverse model can be approximated with deep learning models~\cite{polydoros2015real} or Gaussian process models~\cite{romeres2019derivative}. 
Additionally, data-enabled models can be used to learn the inverse output-to-input map from input-output data ~\cite{yan2022precision}.
Thus, the proposed o-MPPI can use a wide range of inverse modeling tools to find the inputs associated with the sampled output trajectories.

\section{Proposed Framework}
\label{sec_proposed_o_mppi}
The proposed output-sampling-based MPPI (o-MPPI) approach is summarized in Algorithm~\ref{algo_generic_o_MPPI}. Essentially, there are four steps (detailed below): (i)~selection of the cost function; (ii)~sampling in the output space; (iii)~inversion to find the associated inputs; and (iv) weighted selection of the input. 

\begin{algorithm}[]
    \caption{o-MPPI (red is the difference from standard MPPI)}
    \label{algo_generic_o_MPPI}
    \begin{algorithmic}[1]
    \STATE \textbf{Given: }Number of rollouts \& time steps $M,N$; Cost function; Temperature parameter $\lambda$; {\color{red}Inverse model $G^{-1}$};
    \WHILE{Task is not done}
        \STATE $X_k \leftarrow$ state estimate
        \FOR{$m\rightarrow 0$ to $M-1$ in parallel}
        \STATE $Y_0^m \leftarrow h(X_k)$, $S^m \leftarrow 0$
        \STATE [\textbf{Sampling}] {\color{red}Sample the $m^{th}$ trajectory rollout $Y^m$}
        \STATE [\textbf{Inverse}] {\color{red}Compute the corresponding inverse input $u^m$ and states $X^m$ from the inverse model $G^{-1}$}
        \STATE Calculate the trajectory cost $S^m$ based on the cost function and the $m^{th}$ rollout  $(u^m,X^m,Y^m)$
        \ENDFOR
        \STATE [\textbf{Weighting}] For $m=1,2,\dots, M$, compute the normalized weights $\{w_m\}$ based on the trajectory costs $\{S^m\}$ and the selected temperature parameter $\lambda$ 
        \STATE Obtain the weighted average $u=\sum_{m=0}^M w_m u^m$ 
        \STATE Apply the first entry of $u$ to the system
        \STATE $k\leftarrow k+1$
    \ENDWHILE
    \end{algorithmic} 
\end{algorithm}

As in standard MPPI, a cost function is used to specify the desirability of a specific input-state-output rollout. 
In particular, consider the optimization 
\begin{equation}
\min_u J(u) = 
\phi(X_{N-1})+\sum_{k=0}^{N-1}q(X_k)+\frac{1}{2}u^T_kRu_k
\label{eq_predicted_costs}
\end{equation}
subject to system dynamics 
\begin{align}
X_{k+1}&=f(X_k,u_k), Y_k = h(X_k),
\end{align}
where $f$ represents  the forward dynamics, $h$ maps the state $X_k$ to the output $Y_k$ at time step $k$, and the output trajectory is given by $\mathbf{Y}\triangleq \begin{bmatrix}
h(X_0)&h(X_1)&\dots&h(X_{N-1})    
\end{bmatrix}$. In the above optimization, $q(\cdot)$ is the running cost, $R$ is the weight matrix of the input energy cost, and the terminal cost is $\phi(\cdot)$ in Eq.~\eqref{eq_predicted_costs}.
Additionally,  there are constraints on the output trajectory to lie in an   acceptable  region, i.e., 
$\mathbf{Y} \in \cal{Y}$.

\begin{figure}[!ht]
\centering
\includegraphics[width=\columnwidth]{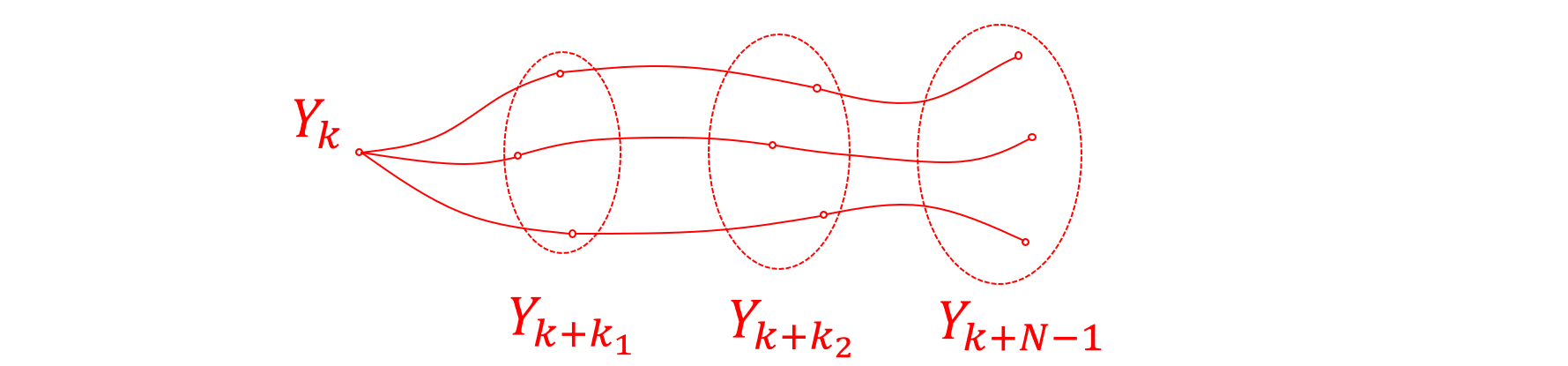}
\caption{A generic way of sampling trajectory rollouts based on fitted smooth curves specified by waypoints that are selected from the regions of interest (dashed ellipsoids).
}
\vspace{-0.1in}
\label{fig_output_rollout}
\end{figure}

The output sampling can be generated from any trajectory planning method, e.g., parametric ones like spline or bezier curves by specifying the waypoints as in Fig.~\ref{fig_output_rollout}, using optimization methods such as $k^{th}$-order constrained path optimization (KOMO)~\cite{toussaint2017tutorial}, via-point-based Stochastic Trajectory Optimization (VP-STO)~\cite{jankowski2023vp}, or from a data-based planner~\cite{arslan2015machine,qureshi2019motion,johnson2020dynamically,wang2020neural}.
For each sampled output, inverse maps are used to find the corresponding input that yields tracking of the output. 

\vspace{-0.1in}

The optimized input is obtained as the weighted sum of all the input rollouts as in standard MPPI, i.e.,
\begin{equation}
\sum_{m=0}^Mw_mu^m=u_{\text{mean}} + \sum_{m=0}^Mw_m\epsilon^m,
\label{eq_average_selection}
\end{equation}
where $u^m=u_{\text{mean}}+\epsilon^m$ and $\{w_m\}$ for $m=1,2,\dots,M$ are are normalized weights, i.e., $\sum_{m=1}^Mw_m=1$.

\section{Application of o-MPPI to Experimental setup}
The system for evaluating the proposed o-MPPI is set up to mimic a dynamic autonomous driving scenario with moving obstacles. 

\subsection{System description}
\begin{figure}[!ht]
\centering
\includegraphics[width=\columnwidth]{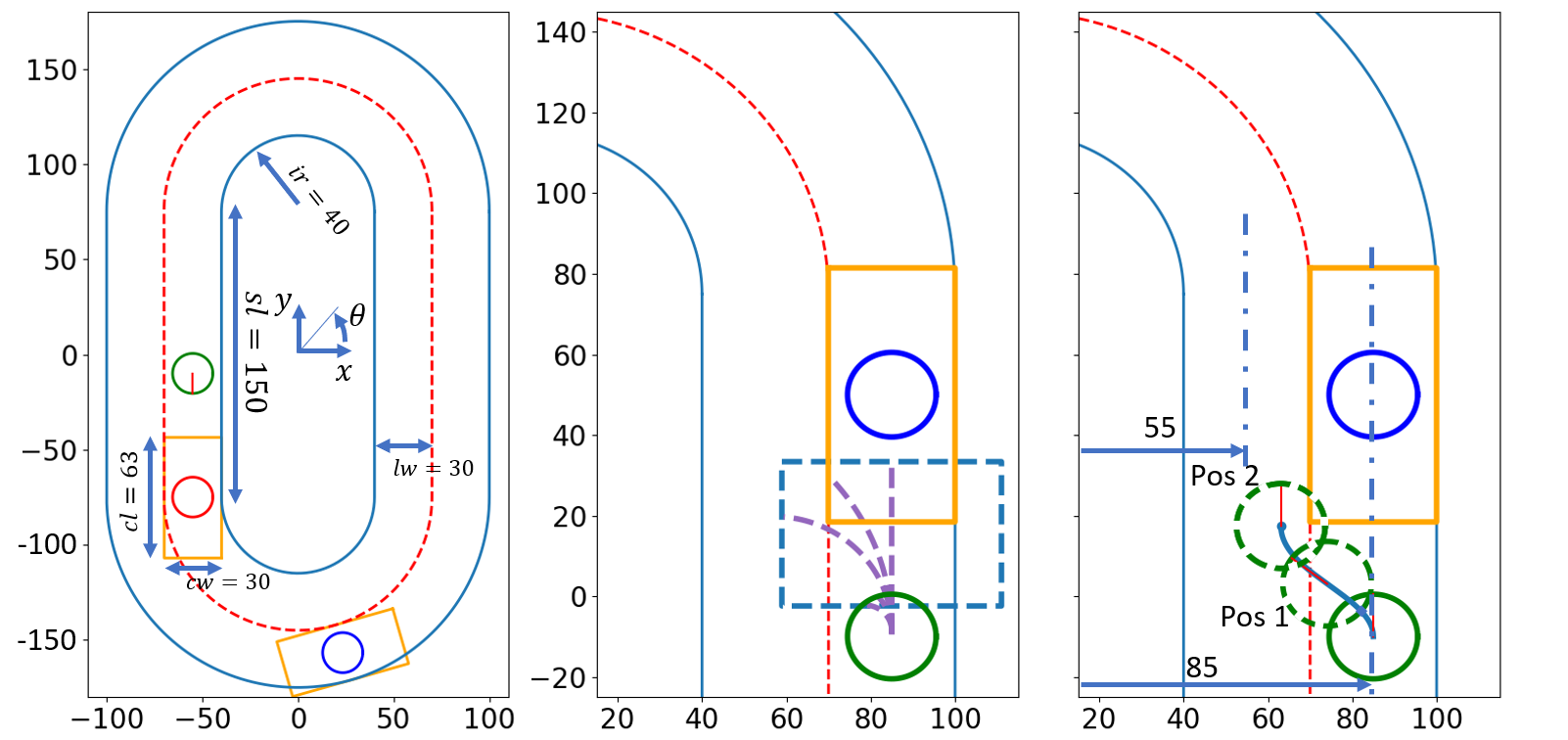}
\caption{The left plot is the schematic drawing of the track. 
The middle plot illustrates an example region of interest (dashed blue rectangle) from which the output waypoint is sampled. The right plot shows an example output rollout where the solid circles indicate the current position and dashed circles indicate future positions in the rollout. The bright yellow rectangles indicate the collision regions.
}
\vspace{-0.2in}
\label{fig_rollout_compute_and_sampled_rec}
\end{figure}

In the experiment, the goal is for a fast bot (green) to move as close as possible to its desired speed of 20 cm/s while maneuvering around slower bots that can be considered as moving obstacles at the same time. The ability for the fast bot to maintain its desired speed and overtake the slower bots, as illustrated in Fig.~\ref{fig_rollout_compute_and_sampled_rec},  are used to quantitatively compare the performance of the standard MPPI and the proposed o-MPPI. An example simulation run is shown in the left plot in Fig.~\ref{fig_rollout_compute_and_sampled_rec}.
The slower bots are moving at constant speeds of 10 cm/s (blue) and 12 cm/s (red). All bots are driving in the counter-clockwise direction in ellipsoidal tracks with dimensions tabulated in Table~\ref{tab_turtle_race_info} and illustrated in Fig.~\ref{fig_rollout_compute_and_sampled_rec}, similar to the shape of the track in~\cite{williams2018information}.

\vspace{-0.1in}
\begin{table}[!ht]
\scriptsize
\renewcommand{\arraystretch}{1.3}
\centering
\caption{Turtlebot, and specifications of track shown in Fig.~\ref{fig_rollout_compute_and_sampled_rec}.
}
\vspace{-0.1in}
\begin{tabular}{|c|c|c|c|c|c|}
\hline
bot turning circle radius (cm)& 10.5 & lane width $lw$ (cm) & 30\\
\hline
max vel. $\Bar{v}$ (cm/s)& 22  &straight lines $sl$ (cm) & 150\\
\hline
max angular vel. $\Bar{\omega}$ (rad/s)& 2.8 & inner radius $ir$ (cm)&40  \\
\hline
collision area width $cw$ (cm)& 30& collision area length $cl$ (cm)& 63 \\
\hline
\end{tabular}
\label{tab_turtle_race_info}
\end{table}   

\begin{figure}[!ht]
\centering
\includegraphics[width=0.8\columnwidth]{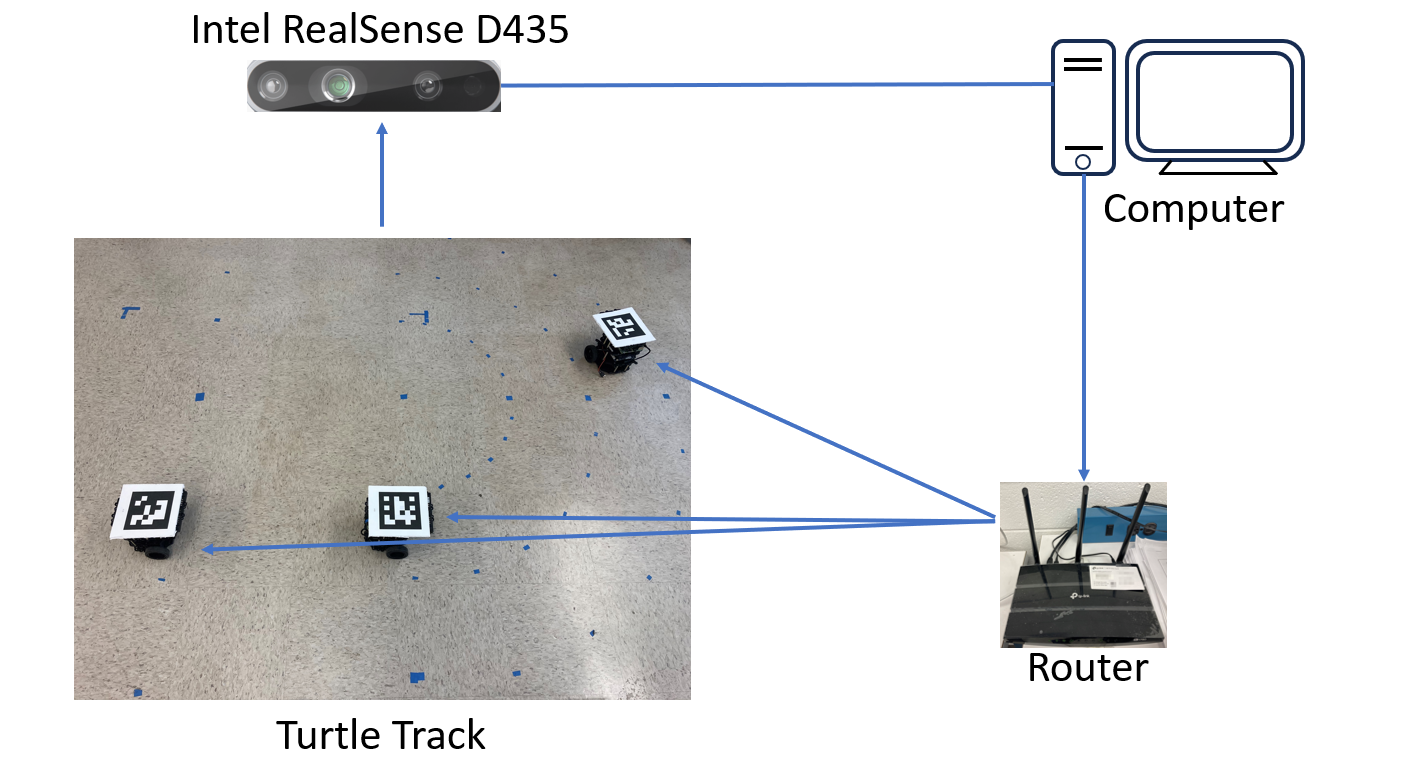}
\caption{The positions of the bots (TurtleBot3 Burger) are estimated from images taken with an  Intel RealSense D435 camera. 
The control algorithms are run on a computer and the control commands are sent to the robots via a wireless router. 
In the Turtle Track photo, the blue tapes are used to indicate the track boundaries.
}
\label{fig_TurtleBot3 Burger_specifc}
\end{figure}

\vspace{-0.2in}
\subsection{Standard MPPI}
\subsubsection{Forward model $F$}
 The bot (TurtleBot3 Burger) state $X$ at time step $k$ is defined as $X_k\triangleq \begin{bmatrix} x_k&y_k&\theta_k&v_k&\omega_k 
\end{bmatrix}$. $(x_k,y_k)$ is the position pair and $\theta_k$ is the orientation with respect to the Turtle Track coordinates, as in Fig.~\ref{fig_rollout_compute_and_sampled_rec}. $v_k$ is the forward velocity and $\omega_k$ is the angular velocity. The input $u$ to the system is $u_k\triangleq \begin{bmatrix}
v^{\text{des}}_k & \omega^{\text{des}}_k
\end{bmatrix}$ where $v^{\text{des}}_k$ is the desired forward velocity and $\omega^{\text{des}}_k$ is the desired angular velocity, and the system output is $Y_k\triangleq \begin{bmatrix}
x_k&y_k
\end{bmatrix}$.

The forward bot dynamics $\Dot{X}=f(X,u)$ is approximated by the forward Euler discretization $X_{k+1}=\mathbf{F}(X_k,u_k)$, 
\begin{align}
x_{k+1} &= x_k + v_k\cos{(\theta_k)}*\Delta t \label{eq_f_1}\\
y_{k+1} &= y_k + v_k\sin{(\theta_k)}*\Delta t\\
\theta_{k+1} &= \theta_k + \omega_k *\Delta t
\\
v_{k+1} &= \text{sat}_{\Bar{v}}(v_k + \alpha(v_k^{\text{des}}-v_k)*\Delta t)\label{eq_f_4}\\
\omega_{k+1} &= \text{sat}_{\Bar{\omega}}(\omega_k + \alpha(\omega_k^{\text{des}}-\omega_k)*\Delta t), 
\label{eq_f_5}
\end{align}
where the sampling time is $\Delta t = 0.04$~s, $\text{sat}_{\beta}(x)= \beta\text{sign}(x)$ if $\text{abs}(x)>\beta$ otherwise $\text{sat}_{\beta}(x)= x$ is a saturation function 
used to bound with the magnitudes of the speed and angular velocity with $\Bar{v}$ and   $\Bar{\omega}$ defined in Table~\ref{tab_turtle_race_info}, 
and the settling-time parameter $\alpha=4/0.35$ is estimated from the measured step response.

\vspace{0.001in}
\subsubsection{Cost function selection}
\label{subsubec_cost_func}
The cost function $J$ in Eq.~\eqref{eq_predicted_costs} is designed to have only the running cost function $q(\cdot)$,
with zero input energy cost ($R$=0), 
similar to the cost functions used in autonomous driving in~\cite{williams2018information,yin2023risk}.
The running cost function is designed to have three components, i.e., $ q(\cdot)\triangleq c_l(\cdot)+c_v(\cdot)+c_c(\cdot)$. The first cost  $c_l(\cdot)$, designed for lane keeping, is the product of the squared values of the bot's distance to the center lines of the two lanes, as shown in the right plot in Fig.~\ref{fig_rollout_compute_and_sampled_rec}. Formally, 
\begin{equation}
c_l(X_k) = w_0(r_k-55)^2(r_k-85)^2 + w_1 \gamma_{OT}(x_k,y_k),
\label{eq_lane_keeping_cost}
\end{equation}
where $w_0 = 0.001, w_1 = 600$, $\gamma_{OT}(x_k,y_k) = 1$ if 
$(x_k,y_k)$ is outside of the designed track shown in Fig.~\ref{fig_rollout_compute_and_sampled_rec} and zero otherwise, $r_k=\sqrt{x^2_k+z^2_k}$, and 
$z_k=0 ~\text{if abs}(y_k)<75.0 $ and 
$z_k= y_k-75.0*\text{sign}(y_k) $ otherwise. 
The second cost 
$c_v(\cdot)$, designed for driving the bot at a speed close to the desired speed of $20$~cm/s,  is the squared of the speed difference between the current speed $v_k$ and the desired one, 
\begin{equation}
c_v(X_k) = w_1(v_k - 20)^2, \qquad w_1=0.4.
\label{eq_speed_cost}
\end{equation}
The third cost 
$c_c(\cdot)$,  designed to avoid collisions with the slower bots, is set to a large value $500$ if $(x_k,y_k)$ is inside the (extended) collision region at time step $k$
\begin{equation}
c_c(X_k,\text{agent}_k)=
\begin{cases}
500 & \text{proj}_f<(0.5cl+cr)\\
&  \text{and } \text{proj}_l<0.5cw,\\
0 & \text{otherwise}
\end{cases}
\label{eq_collide_cost}
\end{equation}
where $cr$ is the turning circle radius, and $cl, cw$ are the collision region length and width, respectively, as seen in Table~\ref{tab_turtle_race_info}. The computation of the projection distance $\text{proj}_f$ and $\text{proj}_l$ are described below
\begin{align*}
\text{proj}_f &= \text{abs}\left(\mathbf{1}_{f}
[\text{dis}_x,\text{dis}_y]^T\right),
\text{proj}_l = \text{abs}\left(\mathbf{1}_{l}
[\text{dis}_x,\text{dis}_y]^T\right)
\end{align*}
where
$\text{dis}_x
=x_{\text{agent}}-x_k,
\text{dis}_y
=y_{\text{agent}}-y_k$, and
\begin{align*}
&\mathbf{1}_{f}=
[\cos{(\theta_{\text{agent}})},\sin{(\theta_{\text{agent}})}],\\
&\mathbf{1}_{l}=
[\cos{(\theta_{\text{agent}}-\frac{\pi}{2})},\sin{(\theta_{\text{agent}}-\frac{\pi}{2})}]. 
\end{align*}
\subsubsection{MPPI algorithm}
\label{subsubsec_mppi_params}
The standard MPPI~\cite{williams2018information}
algorithm~ is summarized in Algorithm~\ref{algo_standard_mppi} (blue parts indicate differences from the proposed o-MPPI) with 
the temperature parameter selected as $\lambda=2.0$ and the covariance matrix is selected as  $\Sigma_{\epsilon}=\text{diag}([4.0,1.0])$ to enable sufficient exploration. 
In general, an additional cost can be added to the cost function for input deviation as $c(\cdot)=\sum_{k=0}^{N-1}\gamma u^T_{k}\Sigma_{\epsilon}^{-1}\epsilon_{k}$~\cite{williams2018information}.
However, the hyperparameter $\gamma$ is set to zero to promote exploration in the dynamic environment with moving obstacles. The mean of the input distribution $u_{\text{mean}}$ in Eq.~\eqref{eq_average_selection} and the prediction horizon $T$ are varied to illustrate their impact on MPPI effectiveness in the results and discussion Section~\ref{sec_results_discuss}.

\subsubsection{Weighting}
Given the trajectory costs $S^m$ for all the rollouts ($0 \le m \le M-1$) based on the cost function,  the weight $w^m$ associated with each rollout ($0 \le m \le M-1$) is computed as 
$w^m=\frac{1}{\eta}(S^m-\beta)$, 
where $\beta$ is the minimum cost 
of the rollouts 
$\beta = \min_m{[S_m]}$
and the normalizing factor $\eta$ is 
$\eta = \sum_{m=0}^{M-1} \exp(-\frac{1}{\lambda}(S_m-\beta))$.

\begin{algorithm}[]
    \caption{Standard MPPI from \cite{williams2017model}}
    \label{algo_standard_mppi}
    \begin{algorithmic}[1]
    \STATE \textbf{Given: }Number of rollouts \& time steps $M,N$; Cost function; Temperature parameter $\lambda$; {\color{blue}Forward model $F$};
    \STATE Control hyperparameters: $\Sigma_{\epsilon},\phi,q,{\color{blue}\gamma\in [0,1]}$ 
    \STATE $(u_0,u_1,..u_{N-1})$ \text{Initial control sequence}
    \WHILE{task not completed}
        \STATE $X_k \leftarrow$ state estimate
        \FOR{$m\rightarrow 0$ to $M-1$ in parallel}
        \STATE $X_0^m \leftarrow X_k$, $S^m \leftarrow 0$
        \STATE [\textbf{Sampling}] {\color{blue}Sample the $m^{th}$ input perturbation rollout $\mathcal{E}^m=\{\epsilon^m_0,\epsilon^m_1,\dots,\epsilon^m_{N-1}\}$}
        \STATE [\textbf{Forward}] {\color{blue}Compute the trajectory rollout $y^m$ from the forward model $F$ and the perturbed input $u^m_{\epsilon}$ computed from $\mathcal{E}^m$ as in Eq.~\eqref{eq_average_selection}}
        \STATE Calculate the trajectory cost $S^m$ based on the cost function and the $m^{th}$ rollout  $(u_m,X^m,y_m)$ 
        \ENDFOR
        \STATE [\textbf{Weighting}] For $m=1,2,\dots, M$, compute the normalized weights $\{w_m\}$ based on the trajectory costs $\{S^m\}$ and the selected temperature parameter $\lambda$ 
        \STATE Obtain the weighted average $u=\sum_{m=0}^M w_m u^m$ 
        \STATE Apply the first entry of $u$ to the system
        \STATE {\color{blue}Warm start~\cite{williams2017model} and $u_{N-1}\leftarrow \text{Initialize}(u_{N-1})$ }
        \STATE $k\leftarrow k+1$
    \ENDWHILE
    \end{algorithmic} 
\end{algorithm}

\subsection{o-MPPI}
The proposed o-MPPI (see Aglorithm~\ref{algo_generic_o_MPPI}) differs from the standard MPPI in the sampling of the output and the use of inverse to find the input -- these are described below. The proposed o-MPPI uses the same temperature parameter $\lambda$, cost function and weighting to determine the input as the standard MPPI case in Sections~\ref{subsubec_cost_func} and \ref{subsubsec_mppi_params}.

\vspace{0.1in}
\subsubsection{Sampling the trajectory rollout}
The generic way in Fig.~\ref{fig_output_rollout} is used to sample the trajectory rollouts in this paper with only two waypoints: the initial output and the final output. Specifically, a rectangle-shape region of interest for the endpoint $Y_{k+N-1}=\begin{bmatrix}x_{k+N-1}&y_{k+N-1}\end{bmatrix}$ is determined based on the allowed magnitudes of the inputs $u$ and the prediction horizon $N$. Then, the $m^{th}$ sampled trajectory rollout is generated by fitting cubic splines between the current output and the $m^{th}$ final output $Y^m_{k+N-1}=(x_e,y_e)$ sampled inside the region of interest, as seen in the right plot in Fig.~\ref{fig_rollout_compute_and_sampled_rec}. 
Specifically, the $m^{th}$ rollout expression $t\in[0,T]$ (for the output $x$) is selected as a cubic spine as in~\cite{zhang1997splines} 
\begin{equation}
x^m_d(t) = a_0t+a_1t+a_2t^2+a_3t^3,
\end{equation}
where the coefficients $a_i$ for $i=1,\dots,4$ can be determined from the boundary conditions 
\begin{equation}
\begin{aligned} 
x(0) &= x_k, \\
x(T) &= x^m_e, \Dot{x}(0) = v^m_e\cos(\theta^m_e),
\Dot{x}(T) = v^m_e\sin(\theta^m_e),
\end{aligned} 
\nonumber 
\label{x_rollout_eq}
\end{equation}
where the endpoint orientation $\theta^m_e$ is designed to be aligned with the road direction and the endpoint forward speed $v^m_e$ is the travel distance divided by the prediction horizon $T$, i.e.,
$v_e = \sqrt{(x_e-x_k)^2+(y_e-y_k)^2}/T$.
The output 
$y^m_d(t)$ is generated in a similar manner with the selected final waypoint output $y^m_e$.

\vspace{0.1in}
\subsubsection{Inverse model $G^{-1}$}
Given a differentiable output trajectory $x_d(t),y_d(t)$ with $t\in [0,T]$ and starting state $X_k$, i.e., $x_d(0) = x_k$, $\Dot{x}_d(0) = v_k\cos{(\theta_k)}$, $y_d(0) = y_k$, $\Dot{y}_d(0) = v_k\sin{(\theta_k)}$, the inverse input  can be obtained at time steps $k+j$ with $j=0,1,\dots,N-1$ from the forward dynamics in Eq.~\eqref{eq_f_4} to \eqref{eq_f_5},  as:
\begin{equation}
\begin{aligned}
v^{\text{des}}_{k+j} &= \frac{v_{p,k+j+1}-v_{p,k+j}}{\alpha\Delta t}+v_{p,k+j}\\
\omega^{\text{des}}_{k+j} &= \frac{\omega_{p,k+j+1}-\omega_{p,k+j}}{\alpha\Delta t}+\omega_{p,k+j}  ,  
\end{aligned}
\label{inv_input_eq_L}
\end{equation} 
where
\begin{align}
v_{p,k+j} &= \sqrt{\Dot{x}^2_d((k+j-k)\Delta t)+\Dot{y}^2_d((k+j-k)\Delta t)},
\\    
\omega_{p,k+j} &=
\begin{cases}
(\theta_{p,k+j}-\theta_{p,k+j-1})/{\Delta t} & j\ge 0
\\
\omega_k & j = 0
\end{cases}\\
\theta_{p,k+j} &=\arctan\left({\frac{\Dot{y}_d((k+j-k)\Delta t)}{\Dot{x}_d((k+j-k)\Delta t)}}\right),
\end{align}
$v_{p,k+j},\omega_{p,s},\theta_{d,s}$ are the planned speed, angular velocity, and the orientation at time step $s$, respectively.  Saturation is not considered when computing the inverse input in Eq.~\eqref{inv_input_eq_L}.  However, if the final optimal input is large (for either o-MPPI or standard MPPI), then it is saturated in the experimental system and bounded in simulations by the saturation function in Eqs.~\eqref{eq_f_4}- \eqref{eq_f_5}. 

\section{Results and discussion}
\label{sec_results_discuss}
The ability to avoid and successfully overtake moving obstacles in dynamic environments are comparatively evaluated below for the proposed o-MPPI and the standard MPPI. The evaluation starts with the study of a single dynamic obstacle, which is followed by the case with multiple obstacles.
\subsection{Case with a single dynamic obstacle}
\label{subsec_overtake_case_study}
For comparative evaluation with a single obstacle, three MPPI and one o-MPPI cases are simulated with the following conditions. 
\begin{itemize}
\item {\bf{Case 1 o-MPPI:}~} The proposed o-MPPI with prediction horizon $T=2.0$~s.\label{case_study_enum_1}
\item {\bf{Case 2 Small-horizon standard MPPI:} ~} Standard MPPI with prediction horizon $T=2.0$~s, i.e., prediction steps $N=50$  
with sampling time period $\Delta t = 0.04$~s.
The nominal speed is $15$ cm/s, i.e., with  
initial control sequence 
(in Algorithm~\ref{algo_standard_mppi}) 
$(u_0,u_1,..u_{N-1}) \equiv [15 \text{ cm/s},0\text{ rad/s}]$;\label{case_study_enum_2}

\item {\bf{Case 3 Large-horizon standard MPPI:} ~} 
Standard MPPI  as in {\bf{Case 2}} except with a larger  prediction horizon $T=8.0$~s and corresponding prediction steps $N=200$;\label{case_study_enum_3}
\item 
{\bf{Case 4 Slow-initial standard MPPI:} ~}
Standard MPPI as in {\bf{Case 2}} except that the initial speed is lower at $10$ cm/s i.e., with  
initial control sequence 
(in Algorithm~\ref{algo_standard_mppi}) 
$(u_0,u_1,..u_{N-1})\equiv [10$~cm/s, $0$~rad/s].
\label{case_study_enum_4}
\end{itemize}
 
\begin{figure}[!ht]
\centering
\includegraphics[width=\columnwidth]{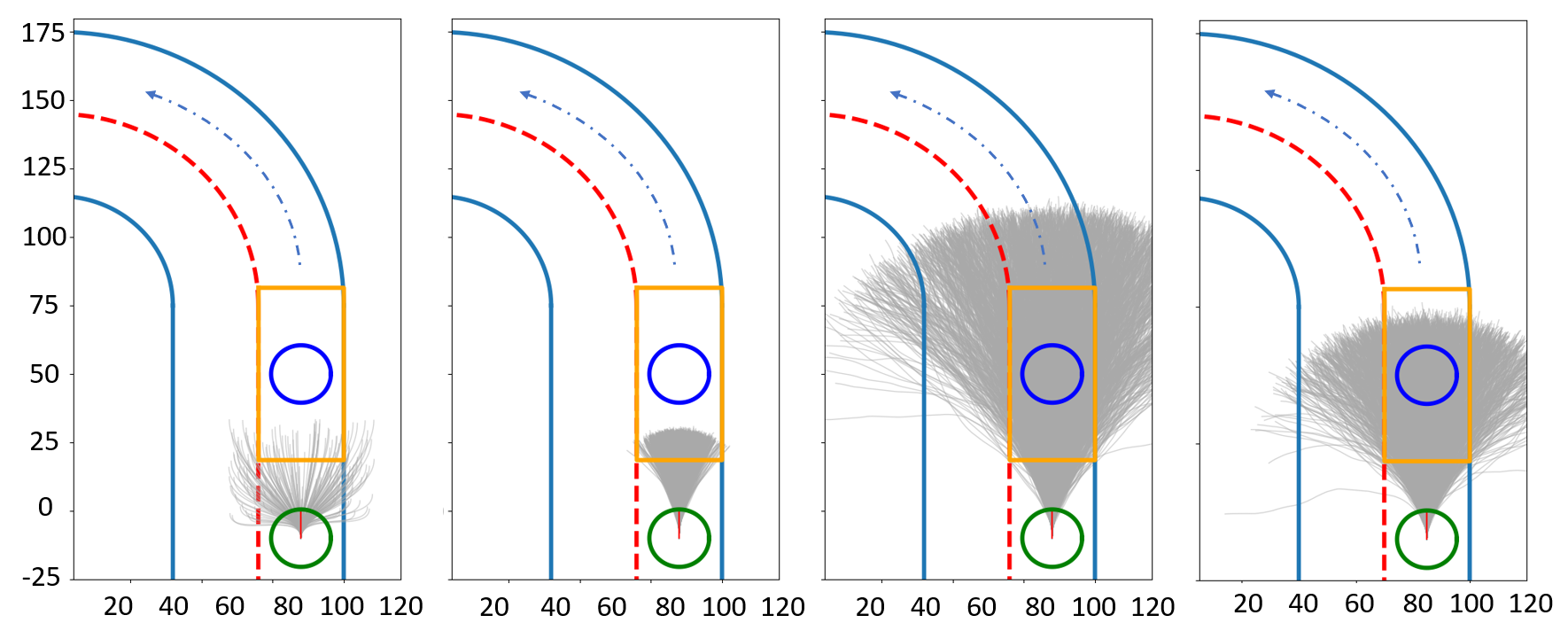}
\caption{Initial rollouts for {\bf{Cases} 1 to 4} (left to right). The green circle represents the controlled bot with the red line indicating its orientation. The blue circle denotes the constant-speed bot of speed $10$ cm/s. The bright yellow rectangle depicts the collision region of the constant-speed bot. The cost function component $c_c$ in Eq.~\eqref{eq_collide_cost}
will be a large value if the controlled bot $(x_k,y_k)$ runs inside the (extended) collision region. Grey lines demonstrate the trajectory rollouts of {\bf{Case 1} to \bf{4}} in Section~\ref{subsec_overtake_case_study}.
}
\label{fig_overtake_case_study}
\end{figure}

\begin{figure}[!ht]
\centering
\includegraphics[width=1\columnwidth]{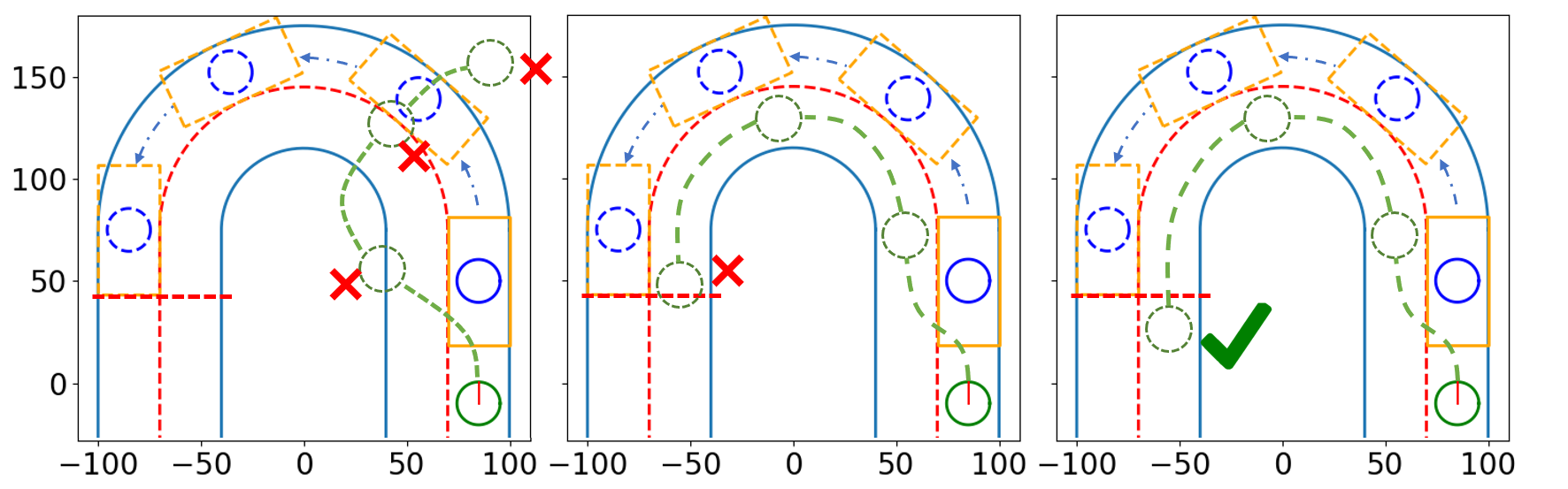}
\caption{
Successful versus unsuccessful overtake, as defined in Section~\ref{subsec_overtake_case_study}.
(Left plot) Unsuccessful overtake where the controlled bot runs outside the track, passes through the moving obstacle, or  drives in the clockwise direction. (Middle plot) Unsuccessful overtake where the final position is not ahead of the moving obstacle's collision region within specified time. (Right plot) Successful overtake that avoids undesirable situations in the left and the middle plots.
}
\label{fig_overtake_success_criteria}
\end{figure}

\vspace{0.1in}
The controlled bot (green in Fig.~\ref{fig_overtake_case_study}) is said to achieve a successful overtake of a moving obstacle (blue circle in Fig.~\ref{fig_overtake_case_study}) if:
(i)~the controlled bot is driving in the counter-clockwise direction all the time,
(ii)~the controlled bot does not run outside of the track or into the collision regions of the moving obstacle, and 
(iii)~within a specified time, the position of the controlled bot is ahead of the slower constant-speed obstacle.
as indicated by the red dashed lines in Fig.~\ref{fig_overtake_success_criteria}. 
For all these cases, the initial position of the moving obstacle is  $(x,y) = (85\text{ cm},50\text{ cm})$ 
on one corner of the track as shown in Fig.~\ref{fig_overtake_success_criteria} 
and it has a constant speed of 
$10$~cm/s. The simulation is run till the moving obstacle reaches the other corner of the track as indicated in Fig.~\ref{fig_overtake_success_criteria}. 
The initial position of the controlled bot is behind the moving obstacle at 
$X_0=\begin{bmatrix}
x_0&y_0&\theta_0&v_0&\omega_0
\end{bmatrix}^T
=\begin{bmatrix}
85\text{ cm}&-10\text{ cm}&\frac{\pi}{2}\text{ rad}&15\text{ cm/s}&0 \text{ rad/s}
\end{bmatrix}^T$.
%
The initial trajectory rollouts for o-MPPI and standard MPPI for the four different cases are visualized in Fig.~\ref{fig_overtake_case_study} for comparative evaluation. Additionally, $100$ repeated simulations were run, and the success rates for the four cases with different numbers of rollouts $M$ are tabulated in Table~\ref{tab_invS_overtake_case_study_different_T_N} and \ref{tab_mppi_overtake_case_study_different_T_N}. 

\vspace{-0.1in}
\subsubsection{Less number of rollouts (efficient exploration)}
The proposed o-MPPI method requires $\mathbf{20}$-times less number of rollouts compared to the standard MPPI, as seen in Tables~\ref{tab_mppi_overtake_case_study_different_T_N} and \ref{tab_invS_overtake_case_study_different_T_N}, 
In particular,  to achieve $100\%$ success rate of overtaking, the standard MPPI requires $M=\mathbf{1000}$ rollouts with a prediction horizon $T=8.0$~s while the o-MPPI method needs $M=\mathbf{50}$ rollouts with a prediction horizon $T=2.0$~s. 
Thus,  the proposed o-MPPI achieves a more efficient exploration compared to standard MPPI.

\vspace{0.1in}
\subsubsection{Smaller prediction horizon}
The o-MPPI method requires $\mathbf{4}$-times smaller prediction horizon compared to the standard MPPI as seen in  Table~\ref{tab_mppi_overtake_case_study_different_T_N} and \ref{tab_invS_overtake_case_study_different_T_N}. In particular,  to achieve $100\%$ success rate in overtaking, the standard MPPI requires a prediction horizon $T=\mathbf{8.0}$~s while the inversion-based sampling method needs only $M=50$ rollouts with a prediction horizon $T=\mathbf{2.0}$~s.

When the standard MPPI method is restricted to a prediction horizon of $2.0$~s, the controlled bot only achieves around $25\%$ success rate, even with a large number of rollouts $M$. This is because o-MPPI has rollouts that run into the other lane as seen in the second left plot in Fig.~\ref{fig_overtake_case_study}. In contrast, rollouts of MPPI (rollout number $M=2000$) with a  prediction horizon $T=2.0$~s) are mostly restricted to within the original lane. Note that the explored area of the standard MPPI increases substantially with a larger prediction horizon of $T=8.0$~s as seen in  Fig.~\ref{fig_overtake_case_study}.

In this sense, the proposed o-MPPI with a smaller prediction horizon requirement is beneficial since requiring a larger prediction horizon might not be feasible due to sensor-range limitations and it can also lead to increased computational load. 

\begin{table}[!t]
\renewcommand{\arraystretch}{1.3}
\centering
\caption{o-MPPI-controlled bot 
success rate ($\%$) 
of overtakes with $T = 2.0$~s 
for {\bf{Case 1}} in Section~\ref{subsec_overtake_case_study}.}
\vspace{-0.1in}
\begin{tabular}{|c|c|c|c|c|c|}
\hline
M & 50 & 100 & 200\\
\hline
success rate& 100&100&100 \\
\hline
\end{tabular}
\label{tab_invS_overtake_case_study_different_T_N}
\end{table}   

\vspace{0.1in}
\subsubsection{MPPI performance is susceptible to the slow initial}
MPPI performance is susceptible to improper initial input sequences, even with sufficient prediction horizon $T=8.0$~s and number of rollouts $M=2000$. For {\bf{Case~4}} with slow-initial speed of $10$ cm/s example, the controlled bot has difficulty overtaking the moving obstacle since the explored area has less overlap with the overtake regions, as seen in Fig.~\ref{fig_overtake_case_study}.
Therefore,  MPPI performance is affected by the initialization strategy and is susceptible to dynamic environments.

\begin{table}[!ht]
\renewcommand{\arraystretch}{1.3}
\centering
\caption{MPPI-controlled bot success rate ($\%$) of overtakes for different prediction horizon $T$ and number of rollouts $M$ for {\bf{Cases 2 to 4}} in Section~\ref{subsec_overtake_case_study}}.
\begin{tabular}{|c|c|c|c|c|c|}
\hline
T & M & succ. rate & T & M & succ. rate \\
\hline
\multicolumn{3}{|c|}{\bf{Case~2}} & \multicolumn{3}{c|}{} \\
\hline
2.0& 50 &28& 4.0& 50& 4\\
2.0& 500&28 & 4.0& 500& 2\\
2.0& 1000&15& 4.0& 1000& 6\\
2.0& 2000& 22& 4.0& 2000& 2\\
\hline
\multicolumn{3}{|c|}{\bf{Case~3}} & \multicolumn{3}{c|}{} \\
\hline
8.0& 50 & 49& 6.0 & 50 & 34\\
8.0& 500 & 75& 6.0& 500 & 52\\
8.0& 1000 & 100& 6.0& 1000 & 89\\
8.0& 2000 & 100& 6.0& 2000 & 97\\
\hline
\multicolumn{3}{|c|}{\bf{Case~4}} & \multicolumn{3}{c|}{} \\
\hline
8.0& 2000 & 1 & &  & \\
\hline
\end{tabular}
\label{tab_mppi_overtake_case_study_different_T_N}
\end{table} 

\subsubsection{Trade-off in sampling time $\Delta t$ selection with MPPI}
\begin{figure}[!ht]
\centering
\includegraphics[width=0.8\columnwidth]{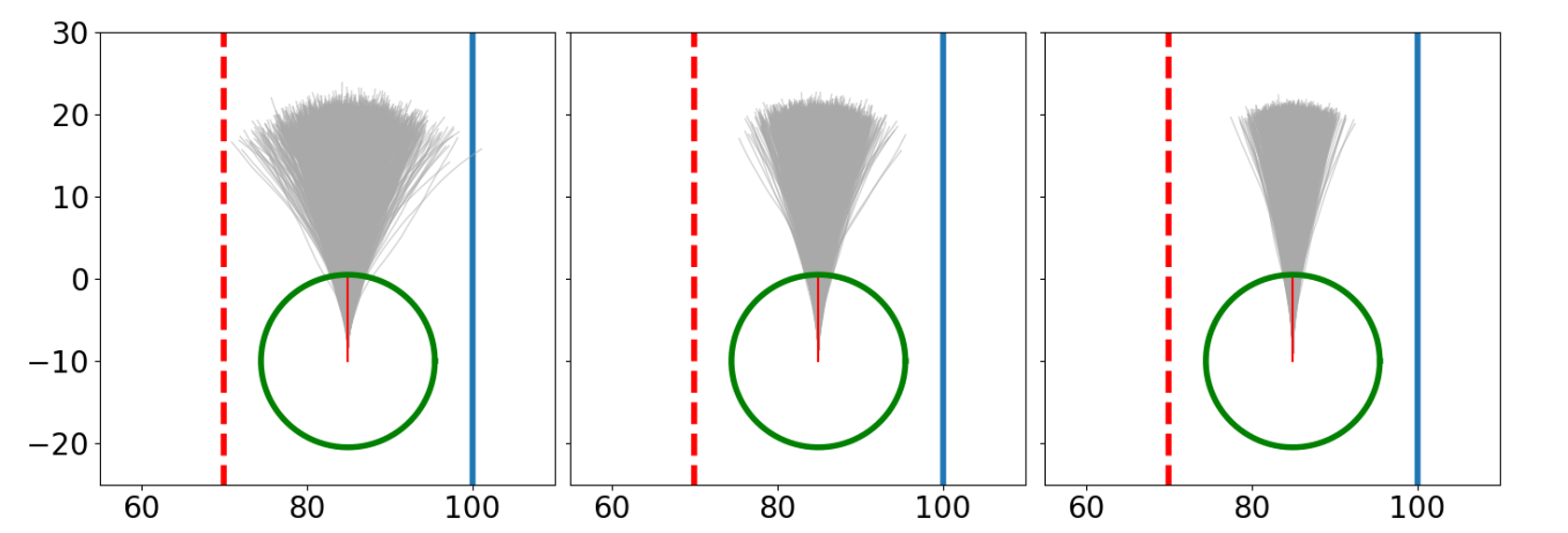}
\caption{The exploration area (black shaded area) is smaller with an increasing sampling rate from 25 Hz (left), 50 Hz (middle) to 100 Hz (right).
All sampling rates have the same prediction horizon of $T=2.0$~s, the same initial state, the same nominal inputs and the same input distribution, and the same number of rollouts $M=2000$.}
\label{fig_control_rate_impact}
\end{figure}

Generally, a higher sampling rate for control is beneficial since this can lead to more precision.
However, a higher rate $\Delta t$ of control also results in a higher dimensional search space for input sampling,
which requires more samples to be drawn for sufficient exploration, as illustrated in Fig.~\ref{fig_control_rate_impact}.  
Less exploration can also result in less success since the best rollout may not be explored when the environment changes too fast or uncertainties are large. 
In contrast, since the o-MPPI trajectory rollouts are sampled from the potential set of waypoints in the output space that only depends on the prediction horizon $T$, the control sampling rate $\Delta t$ does not affect the explored region.

\subsection{Case with multiple dynamic obstacles}
\label{subsec_multiple_obstacles}
\vspace{-0.051in}
For comparative evaluation with multiple obstacles, two moving obstacles are considered as in the left plot of Fig.~\ref{fig_rollout_compute_and_sampled_rec}.
The cost function is kept the same as in the previous Section~\ref{subsec_overtake_case_study} with a single dynamic obstacle.

The first obstacle has a constant speed of $10$~cm/s (red circle) and tracks the inner lane with initial position $(x,y) = (23\text{ cm},-156\text{ cm})$ and the second obstacle has speed $12$~cm/s (blue circle) and tracks the outer lane with initial position $(x,y) = (-55\text{ cm},-75\text{ cm})$.
The initial position of the controlled bot (green) is behind the first moving obstacle and has the initial condition 
$X_0=\begin{bmatrix}
x_0&y_0&\theta_0&v_0&\omega_0
\end{bmatrix}^T
=\begin{bmatrix}
-55\text{ cm}&-10\text{ cm}&-\frac{\pi}{2}\text{ rad}&15\text{ cm/s}&0 \text{ rad/s}
\end{bmatrix}^T$
Comparative simulation and experiments of o-MPPI and standard MPPI were performed with settings as in {\bf{Case 1}} (o-MPPI) and {\bf{Case 3}} (Larger-horizon standard MPPI) proposed in Section~\ref{subsec_overtake_case_study}.

\vspace{0.1in}
\subsubsection{Comparison of two obstacle avoidance}
\begin{figure}[!ht]
\centering
\includegraphics[width=\columnwidth]{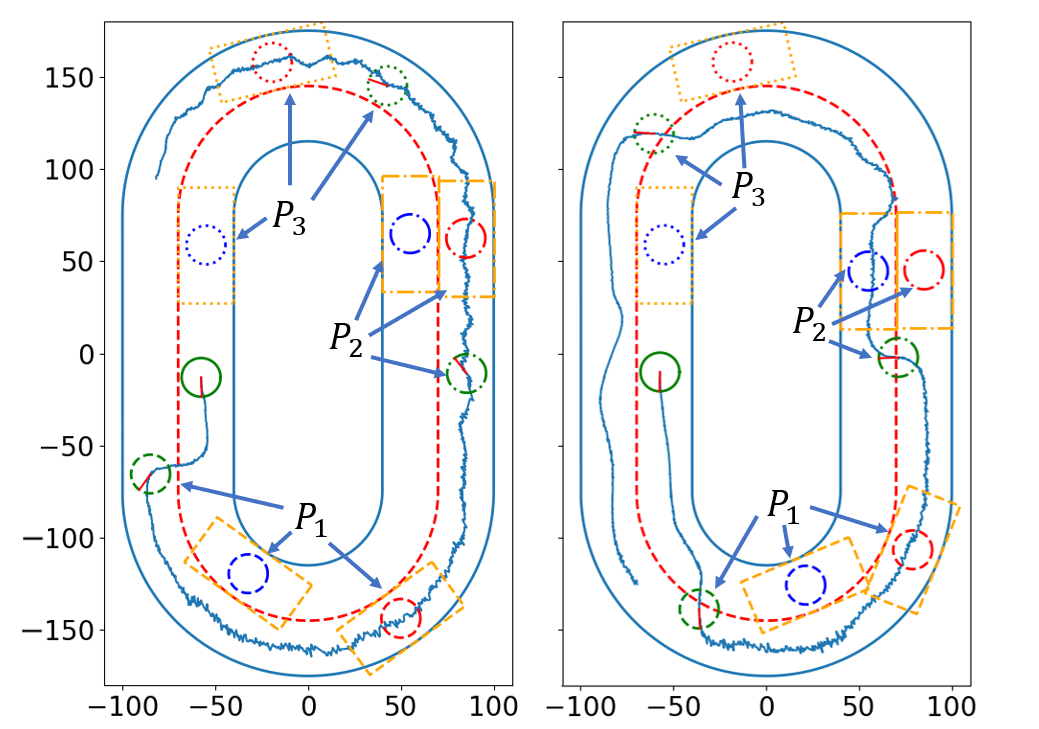}
\caption{
Experimental traces of the MPPI-controlled bot (left) and o-MPPI-controlled bot (right). The o-MPPI-controlled bot could maneuver multiple times (snapshots of positions are indicated as $P_1$ (dashed), $P_2$ (dash-dotted), and $P_3$ (dotted plots)) to overtake two constant-speed moving bots. However, the MPPI-controlled bot got stuck behind the constant-speed moving bot (blue) although it switched to the outer lane initially (at position $P_1$ (dashed plot)) to avoid a single moving bot. Note that the  MPPI-controlled bot moved in a zig-zag manner, which is also seen in Fig.~\ref{fig_snapshot_overlay}. The full experiments can be seen in the accompanying video at {\sffamily{https://youtu.be/snhlZj3l5CE}}.}
\label{fig_trace_plot}
\end{figure}

Only the o-MPPI-controlled bot managed to successfully maneuver around both moving obstacles after slowing down because both lanes were blocked by the two slower constant-speed bots as seen in the trace plots in Fig.~\ref{fig_trace_plot}. 
First, the o-MPPI-controlled bot switched to the inner lane since the inner-lane, constant-speed bot has a relatively faster speed at $12$~cm/s compared to the outer-lane, constant-speed bot moving at $10$~cm/s and second then the o-MPPI-controlled bot switched back to the outer lane once there is sufficient room to drive at a faster speed that is closer to the desired speed $20$~cm/s. In contrast, the standard MPPI-controlled bot got stuck behind the constant-speed bot, as seen from the trace in the upper region of the track in Fig.~\ref{fig_trace_plot}. As shown with {\bf{Case 4}} (slow-initial standard MPPI), with a smaller speed the nominal mean of the MPPI input distribution become smaller, and can lead to lower success rate for overtaking. It is noted that both the o-MPPI-controlled bot and the MPPI-controlled bot were able to successfully maneuver around a single obstacle when the controlled bot is moving at a higher speed as seen in the accompanying video at https://youtu.be/snhlZj3l5CE.

\vspace{0.1in}
\subsubsection{Oscillatory driving versus overtaking}
\begin{figure}[!ht]
\centering
\includegraphics[width=\columnwidth]{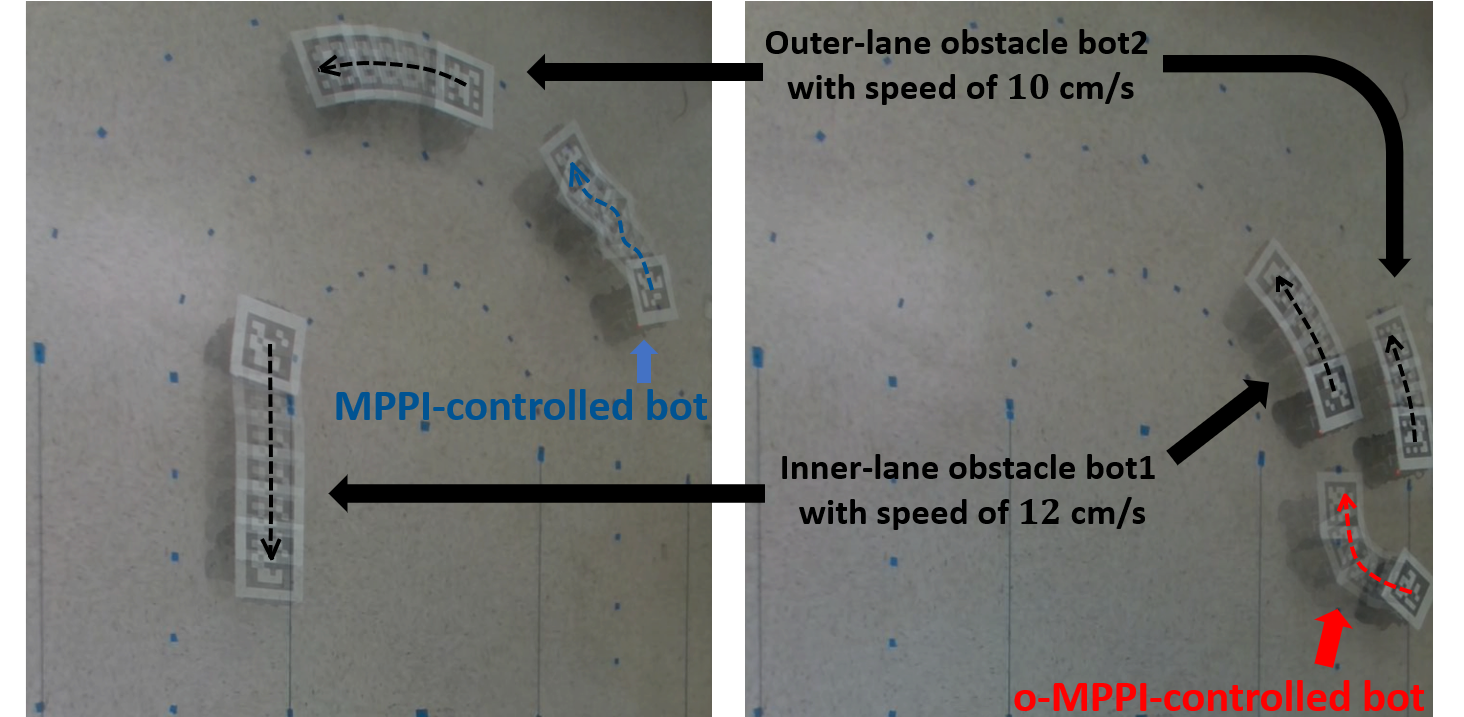}
\caption{
Overlays of experimental images for multiple dynamic obstacles: (i)~(left) the standard MPPI  with a controlled bot that remains stuck and (ii)~(right) the proposed o-MPPI where the controlled bot successfully manages to switch lanes and eventually overtake the slower moving obstacle. The full experiment can be viewed in the accompanying video at {\sffamily{https://youtu.be/snhlZj3l5CE}}.
}
\vspace{-0.1in}
\label{fig_snapshot_overlay}
\end{figure}

Comparison of overlays of the experimental snapshots
in Fig.~\ref{fig_snapshot_overlay} (left) shows that the MPPI-controlled bot can not only get stuck behind the slower moving obstacle bot (with speed $10$~cm/s in the outer lane), but it also tends to have oscillatory driving behavior. This is because the oscillatory motion allows the bot to achieve a relatively faster speed closer to its desired speed in the cost function component $c_v$ in  Eq.~\eqref{eq_speed_cost}.
In contrast, overlays of the o-MPPI-controlled bot in Fig.~\ref{fig_snapshot_overlay} (right) show that it is able to maintain higher speed by successfully switching to the inner lane since the obstacle bot in the inner lane in the front has a higher speed $12$~cm/s compared to the bot in the outer lane with speed $10$~cm/s.

\section{Conclusion and Future works}
This paper proposed an output-sampling-based 
model predictive path integral control
(o-MPPI) to improve the efficiency of standard MPPI. An advantage of the proposed output sampling is that it can leverage the substantial work on trajectory planning in robotics to meet constraints that are often posed in the output space, which in turn improves the efficiency of MPPI. Instead of forward models that map from input to output as in standard MPPI, the proposed o-MPPI uses inverse models to map from the sampled output to inputs. The improved efficiency of the o-MPPI was seen in both the simulation and the experimental results --- o-MPPI required less number of rollouts and a shorter prediction horizon, compared to the standard MPPI. Future works will explore the use of other trajectory-planning methods for output sampling with o-MPPI, as well as the use of data-enabled deep-learning and Gaussian process models for the inverse map. A blending of the input sampling and output sampling can also be considered depending on the type of constraints. 

\addtolength{\textheight}{-13cm}



\bibliographystyle{unsrt}
\bibliography{papers}
\end{document}